\documentclass[%
aip,
 amsmath,amssymb,
 reprint,%
]{revtex4-1}

\usepackage{graphicx}
\usepackage{dcolumn}
\usepackage{bm}

\begin{document}

\title[]{Sensitivity Limit of Nanoscale Phototransistors}

\author{M. Rezaei}
\email{mohsen.rezaei@ieee.org}
\author{M. S. Park}
\author{C. L. Tan}
\author{H. Mohseni}
\email{hmohseni@northwestern.edu}
\affiliation{%
Bio-inspired Sensors and Optoelectronics Laboratory (BISOL), Department of Electrical Engineering and Computer Science, Northwestern University, Evanston, Illinois 60208, USA.
}%


\begin{abstract}
In this paper the optical gain mechanism in phototransistor detectors (PTDs) is explored in low light conditions. An analytical formula is derived for the physical limit on the minimum number of detectable photons for the PTD. This formulation shows that the sensitivity of the PTD, regardless of its material composition, is related to the square root of the normalized total capacitance at the base layer. Since the base total capacitance is directly proportional to the size of the PTD, the formulation shows the scaling effect on the sensitivity of the PTD. We used the extracted formula to study the sensitivity limit of a typical InGaAs/InP heterojunction PTD. Modeling predicts that a PTD with a nanoscale electronic area can reach to a single photon noise equivalent power even at room temperature. The proposed model can also be used to explore the sensitivity and speed of the nanowire-based photodetectors. To the best of our knowledge, this is the first comprehensive study on the sensitivity of the PTD for low light detection.%
%
%
\end{abstract}

\maketitle



\section{introduction}
\begin{figure}
\includegraphics[width=80mm]{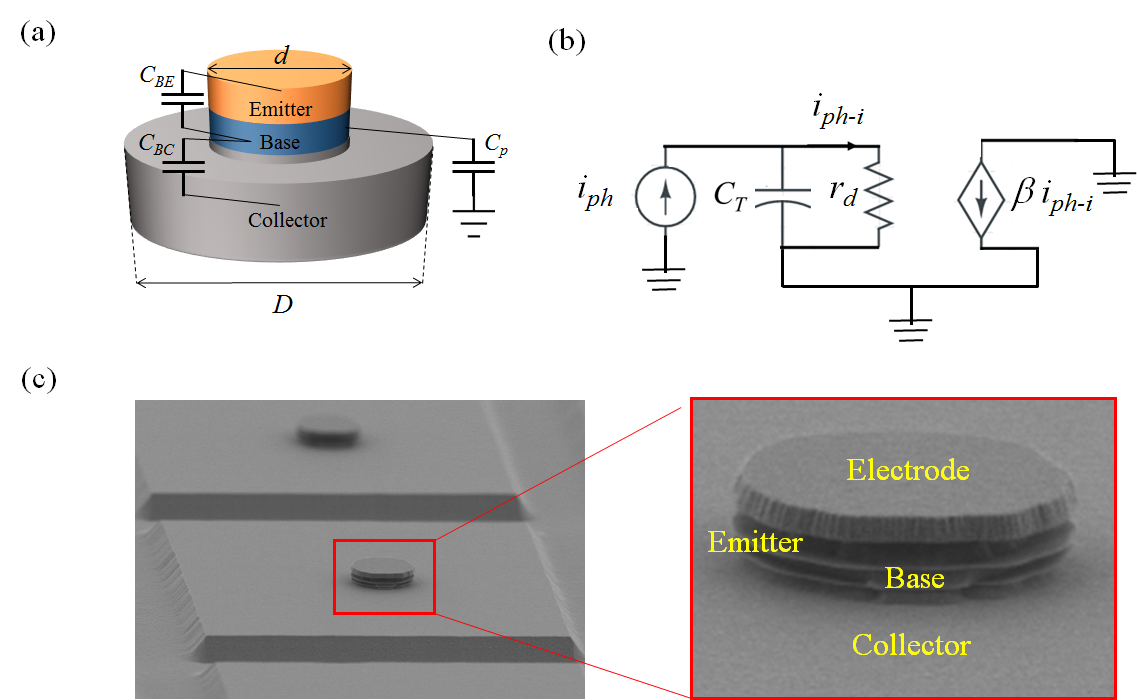}
\caption{Different schematic models for a two-port PTD including (a) physical schematic, (b) low light circuit model for the PTD (c) SEM images of an InGaAs/InP heterojunction phototransistor with different sizes of collector and base.}
\label{fig:model}
\end{figure}

Increasing the sensitivity of infrared astronomical cameras in low light conditions is the key for our further understanding of the universe. Such cameras can also benefit other fields such as medical imaging, light detection and ranging (LiDAR) and quantum computing \cite{Review_SPD}. Here we study the possibility of using phototransistor detectors (PTDs) for such application. To do that we study the relation between the size of PTDs and their speed and sensitivity for low light imaging. It is a well known fact that scaling down the size of transistors enhances their performance \cite{Nano-Wire_PT1,Nano_wire_PT2,Nano1,nano_ZNO,nture_APD}. Specifically nano-wire bipolar junction and field effect transistors showed a promising result for future advancement in electronics and optoelectronics \cite{Nano_GAN,Nano_PT_InAs,Silicon_nano}. However, to the best of our knowledge, the size-dependent performance of such devices has not been analytically explained. Previously, we studied the effect of material composition and defects in PTDs using detailed numerical and analytical modeling, but for a particular device  design \cite{yashar}. Here we present a material-independent holistic view that hows the relationship between PTD geometry, speed, and sensitivity. Our model shows that a PTD with a nanoscale diameter for its electronic part, $d$ , and a diameter of a few microns for the the optical absorption area, $D$, can be used to make a detector array with the noise equivalent power of a single photon (see Fig.\ref{fig:model} for $d$ and $D$).\

PTDs, like other photon detectors, convert the energy of absorbed photons into an electrical signal. In this process, noise is the main obstacle to detecting weak light. There are three main sources of noise in any photon detection system: photon noise, dark noise and read noise. Photon noise is the photon's statistical noise that obeys Poisson statistics. Dark noise is the noise associated with the detector and read noise is the noise of the electronic circuitry that is needed to read the generated electric signal. The lower the overall noise of the detector, the higher its sensitivity  will be. Sensitivity can be simply defined as the minimum number of photons that produce an electric field equal to the total noise. Reducing the temperature of the detector reduces the contribution of its noise on the overall noise level. For detectors without internal gain, e.g., PIN detectors, lowering the temperature beyond a certain level no longer reduces its noise level. The reason is that the sensitivity becomes limited by the read noise. An internal low noise amplifier needs to be added to the detector to address this issue. The read noise contribution will diminish by a factor equal to the value of internal gain.  There are a few mechanisms by which one can add internal gain to the detector. These include avalanche and transistor action. An avalanche photo detector (APD) is a high speed detector that uses avalanche as its internal gain mechanism.  APDs have been widely used to increase the sensitivity of detection systems. Drawbacks of the avalanche mechanism are its excess noise, high voltage operation and  low gain \cite{avalanche}. The voltage of operation for APDs is higher than the operating voltage for standard CMOS electronics, so special circuitry is usually needed to drive them. These drawbacks impose some limitations on using APD for low light detection.\

PTD uses bipolar transistor action for its internal gain. Because of PTD's low voltage operation and high gain, it has been investigated by many researchers for numerous applications, especially in optical communication. Due to the advances in CMOS technology that enabled creating low noise, high speed and low cost amplifiers, attention to the PTD gradually reduced. For ultra-fast applications a PIN photodiode combined with a CMOS read circuit became a better solution. PTD is an inherently slow device and not suitable for telecommunication especially when it is used in the two-port mode. In telecommunication, speed is the most important parameter and usually the power level is far more that what could be considered a low light level\cite{opt_communication,Gokab-main,Arash,Min_Su}. In addition to speed, the other main problem of the PTD is its gain drop in  low power levels, which has imposed huge restrictions on its application. By demonstrating the advance in III-V semiconductor material quality and addressing the gain drop problem \cite{mohsen}, here we show that PTD has  great potential to be used for weak light detection.\

The noise characteristics  of the PTD in the low light regime need to be analyzed for a better understanding of its potential applications. Helme and Houtson have analytically studied the noise and speed of the PTD \cite{PT-analytic}. Recently  Gabrel et. al. also proposed a noise model for pnp PTD \cite{Noise-PT_2014}. In all of the mentioned works the PTD is modeled for high-power and high-speed application. To the best of our knowledge there is no comprehensive study on the PTD for applications close to single photon detection. Our focus in this paper is on the performance analysis of the PTD for low light level detection.\

\section{PTD low light model and sensitivity extraction}
\subsection{circuit model}
Here a low light model is developed to extract the sensitivity of the PTD.  Fig. \ref{fig:model}(a) shows a schematic diagram for a two-port PTD. The device has the diameter of $d$, for emitter and base and a bigger diameter of $D$ for the  collector. A bigger diameter for the collector increases the absorption area and the smaller diameter for the base and the emitter reduces the total capacitance at the base. Part (b) of the figure shows the low light circuit model for the PTD. Part(c) shows SEM images of a PTD with InP/InGaAS/InGaAs structure.  To operate PTD  for low light detection it should have extremely low dark current. For this reason it needs to operate at low temperature.  Since the current levels in the low light condition are extremely low, the series resistances at the base and emitter are being ignored. For a transconductance  amplifier, the input impedance needs to be very low so we can also assume all the capacitances that are connected to the base are parallel.

In the model $C_T$ is the total capacitance at the base which includes the base-emitter ($C_{BE}$), the base-collector ($C_{BC}$) junction capacitance, and all the parasitic capacitance ($C_{p}$) connected to the base. Internal dark current ($I_{d}$) is the base bias current and photocurrent ($i_{ph}$) is the signal.   $\beta$ represents the current gain of the PTD and $i_{ph-i}$ is the base current. There is no assumption of the material system of the PTD, so the model is generalizable for all sorts of PTDs.  External dark current, $I_{d-ext}$, is given by $I_{d-ext}= (\beta+1)I_{d},$ which can be measured from the terminals of the PTD. Throughout this paper the term "dark current" always refers to the  internal dark current. Dark current is not only a source of noise, but also bias current in the current operating regime. Photocurrent, which is the current generated by absorbing photons, is the small signal disturbance for the operation of the PTD. We will show that the higher the dark current, the faster the PTD will be. In the model, $r_d$ is the active mode resistance between the base and emitter and is given by
\begin{equation}
 r_d=\frac{V_t}{I_{d}},%
\end{equation}
where $V_t$ is the thermal voltage that can be expressed as
\begin{equation}
V_t=\frac{KT}{q}.
\end{equation}
In this formula, {\it{K}} is the Boltzmann constant, {\it{T}} is the temperature and {\it{q}} is the elementary charge. For the above circuit, the rise time for a rectangular photocurrent pulse is given by: 
\begin{equation}
t_{rise}=2.2r_d C_T=2.2\frac{V_t C_T}{I_{d}}.%
\label{eq:rise_my}
\end{equation}
In Ref.~\onlinecite{PT-analytic} the rise time is extracted for the PDT as follows:
\begin{eqnarray}
t_{rise}=2.2(\tau _e+\beta R_TC_{BC}+\frac{\beta C_T}{2})+ \nonumber\\
\Bigg(2.2+ln\Bigg\{\frac{I_{d}+
0.1i_{ph}}{I_{d}+0.9i_{ph}}\Bigg\}\Bigg)\frac{V_t}{I_{d}+i_{ph}}C_T.\nonumber\\%
\label{eq:rise_analytic}
\end{eqnarray}
Details of the formula are well explained in the above reference. Comparing Eq.\ref{eq:rise_my} and Eq.\ref{eq:rise_analytic} , we see that for the low light condition, in which the speed dominates with the dark current and junction capacitance, both formulas give the same rise and fall time.\

\subsection{Noise analysis and sensitivity extraction}
Here the noise characteristics of the PTD is studied using the proposed low light model.  A general expression for the signal to noise ratio of a PTD is given by \cite{Memis-noise}

\begin{equation}
SNR^2=\frac{i_{ph-i}^2}{i_{n,ph}^2+i_{n,d}^2+i_{n,r}^2},
\label{SNR}
\end{equation}
where $i_{n,d}$ , $i_{n,ph}$ and $i_{n,r}$ respectively are the dark noise, photon noise and the  effective read noise. $i_{ph-i}$ is the portion of $i_{ph}$ that passes through the dynamic resistance and gets amplified. For a rectangular input light pulse of duration $T$ containing total number of photons of $N$,  peak value of $i_{ph}$ can be written as
\begin{equation}
i_{ph}=\frac{\eta Nq}{T}.
\label{eq:photo-current}
\end{equation}
where $\eta$ is the quantum efficiency. Considering the time response of the PTD, $i_{ph-i}$is given by 

\begin{equation}
i_{ph-i}=(1-e^{-\frac{T}{\tau}})i_{ph}=G\frac{\eta Nq}{T}.
\label{eq:signal-current}
\end{equation}where,  $\tau=r_dC_T$ is the time constant of the PTD. The term $G=1-e^{-g}$ is representing the effect of the limited response time of the PTD. $g=T/\tau$ is the ratio between pulse duration and the time constant of the PTD. In other words, $g$ is the ratio between the measurement speed and the PTD speed.   \

Assuming the shot-noise characteristics for the dark current, $i_{n,d}$  can be expressed as
\begin{equation}
i_{n,d}^2={2qI_d\gamma BW}
\label{eq:dark-noise}
\end{equation}
where $BW$ is the noise bandwidth and $\gamma$ is the fano factor for the shot noise\cite{Shot-noise-Review}. The photon noise can also be written as 
\begin{equation}
i_{n,ph}^2={2qi_{ph-i}BW}
\label{eq:photon-noise}
\end{equation}
As the model shows PTD has an internal pole due to the junction capacitance and the the dynamic resistance. For such a system $BW$ can be approximated by \cite{book}
\begin{equation}
BW=\frac{1}{4\tau}
\end{equation}
Substituting Eq.\ref{eq:signal-current} , Eq.\ref{eq:dark-noise} and Eq. \ref{eq:photon-noise}  in SNR formula in Eq.\ref{SNR} gives
\begin{equation}
SNR=\frac{\eta N G}{\sqrt[]{\frac{g^2\gamma}{2}\frac{C_T}{C_0}+\frac{g}{2}\eta N G+\frac{N_r^2}{G^2\beta^2}}},
\label{eq:SNRR}
\end{equation}
where $C_0$ is defined as the thermal fundamental capacitance  and is expressed by
\begin{equation}
C_0=\frac{q}{V_t}.
\label{eq:C0}
\end{equation} $N_r$ is the number of read-noise electrons. Dividing $N_r$ by the effective gain of the PTD, $G\beta$, gives the  contribution of the read noise on the SNR. 
It is worth mentioning that we can not arbitrarily choose the $g$ to be too small or too big. For a given photon flux, lower values for $g$ increases the read-noise contribution and higher values for that increases the dark-noise contribution of the device.   The SNR formula can be simplified by assuming negligible effect of the read-noise and  $g=1.6$  to 
\begin{equation}
SNR=\frac{\eta N}{\sqrt[]{2\gamma\frac{C_T}{C_0}+\eta N }}.
\label{eq:SNR_simple}
\end{equation}
 In practice, a lot of considerations on the material and the fabrication should be taken into account to make sure that the PTD's gain in ultra-low power range stays above a certain limit \cite{gain_of_HPT} so the PTD can eliminate the effective read noise. 
Solving Eq.\ref{eq:SNR_simple} for $\eta N$ gives
\begin{equation}
\eta N=\frac{1}{2}SNR^2(1+\sqrt[]{1+\frac{8\gamma}{SNR^2}\frac{C_T}{C_0}})
\label{eq:N}
\end{equation}

\section{Results and discussions}

\begin{figure}
\includegraphics[width=80mm]{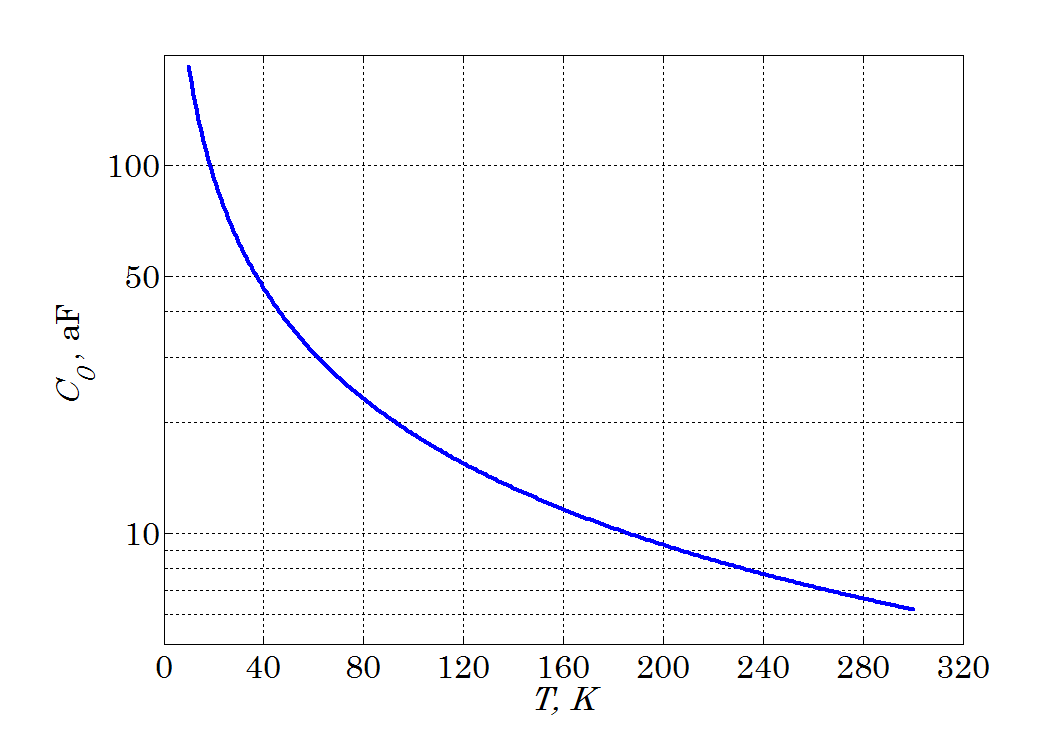}
\caption{\label{fig:C_0} Value of the junction capacitance for a single photon noise equivalent power, $C_0$,  versus temperature (see Eq. \ref{eq:C0}).}
\end{figure}

Here the extracted formula for the sensitivity will be examined for a general PTD. The value for the fundamental thermal capacitance , $C_0$, versus temperature  is illustrated in the Fig. \ref{fig:C_0}. $C_0$ increases from 6.2 attofarad at 300K to almost 186 attofarad at 10K.\ 

\begin{figure}
\includegraphics[width=80mm]{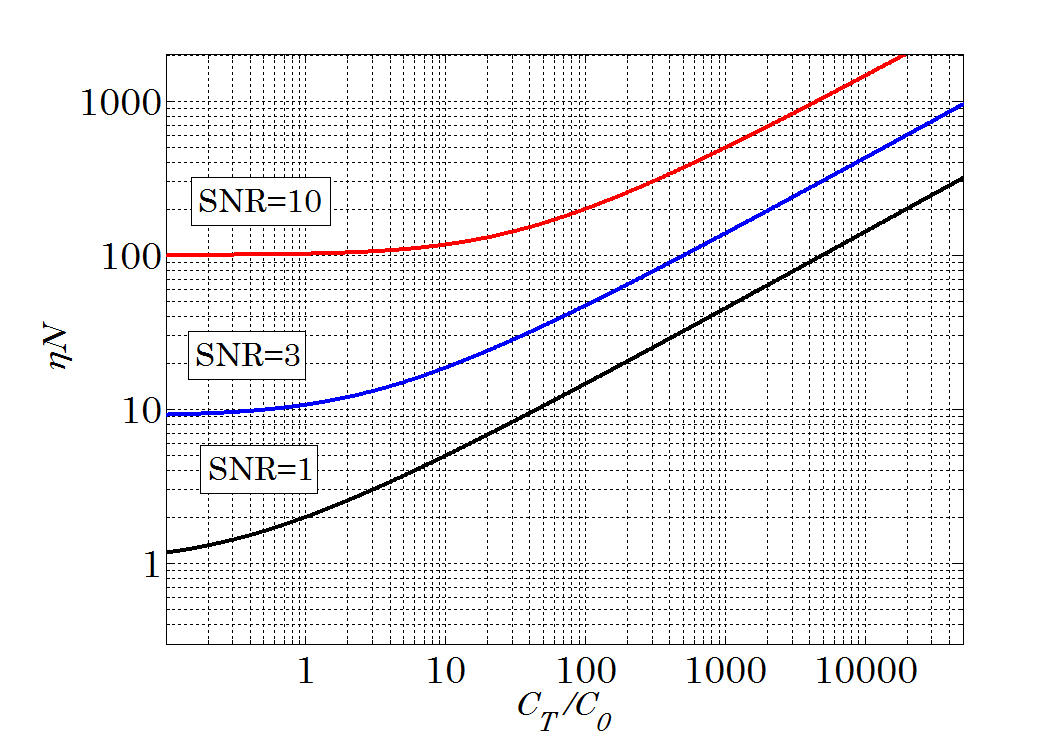}
\caption{\label{fig:sensitivity_vs_cap} Minimum Number of  photons that can be detected with SNRs of 1, 3 and 10 (see Eq. \ref{eq:N}).}
\end{figure}

\begin{figure}
\includegraphics[width=80mm]{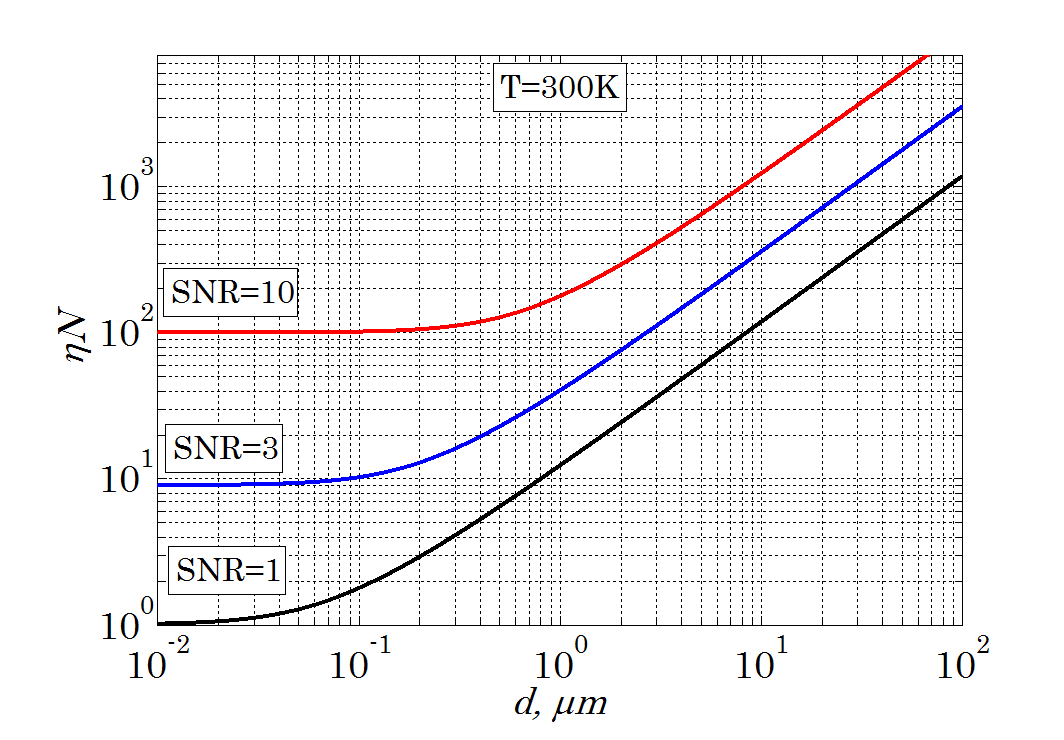}
\caption{\label{fig:number.vs.diameter} Sensitivity of an  InP/InGaAs/InGaAs heterojunction PTD versus its electronic area diameter, \textit{d} (see Ref.\onlinecite{InGaAs_repeat} for the device structure).}
\end{figure}
	The minimum number of photons that create a signal with SNR=1, 3 and 10 versus the normalized junction capacitance is shown in Fig. \ref{fig:sensitivity_vs_cap} (Eq.~\ref{eq:N}). This is clear from the figure that the SNR becomes photon noise limited at low enough normalized junction capacitances.\

To have some estimation of the relation of the size of the PTD  and the minimum number of photons that it can detect, we look at a junction capacitance versus the diameter of the electronic area, $d$. As it is mentioned earlier total capacitance at the base includes two junction capacitances, $C_{BE}$ and $C_{BC}$ and the parasitic capacitance of the base.  In any PTD, the base-emitter junction is forward biased and the base-collector junction is reversed biased.  For a p-n junction with area of A and depletion width of $w_j$  the junction capacitance can be expressed by
\begin{equation}
 C_j=\epsilon_0\epsilon_r\frac{A}{w_j}. 
\end{equation}
In this formula $w_j$ is related to the applied bias voltage and doping concentration of the both p and n sides. This formula ignores the fringe capacitances, and would underestimate the value as diameter goes below $\sim 50 nm$. For a PTD the collector should have much lower doping than the emitter so the depletion width at the base-emitter junction will be smaller than that at the base-collector junction. Therefore, the total capacitance at the junction is mainly determined with base emitter capacitance. As an example, for a short-wave infrared (SWIR) PTD with InP/InGaAs/InGaAs alloy structure and doping concentration of $n10^{16}/p10^{18}/n10^{15}$ the depletion widths for the base-emitter and base-collector junctions at the 1.2V bias voltage, are almost 200nm and 1.5$\mu m$, respectively \cite{InGaAs_repeat} . Fig. \ref{fig:number.vs.diameter} shows the number of detectable photons with a given SNR versus $d$. Without considering parasitic capacitance, this PTD can reach to photon noise limited detection for a single photon at almost 100nm of diameter of the base.  This is clearly much smaller than the diffraction limit of light for the SWIR wavelength. Having a collector diameter as the the optical absorption area which is bigger than the base diameter helps to efficiently absorb the light and use the diffusion to capture the carriers. So the diffusion length of the carriers  sets the limit for the collector diameter.  This diameter, which defines the optical spot size, can be tens of microns for conventional semiconductors. \
\begin{figure}
\includegraphics[width=80mm]{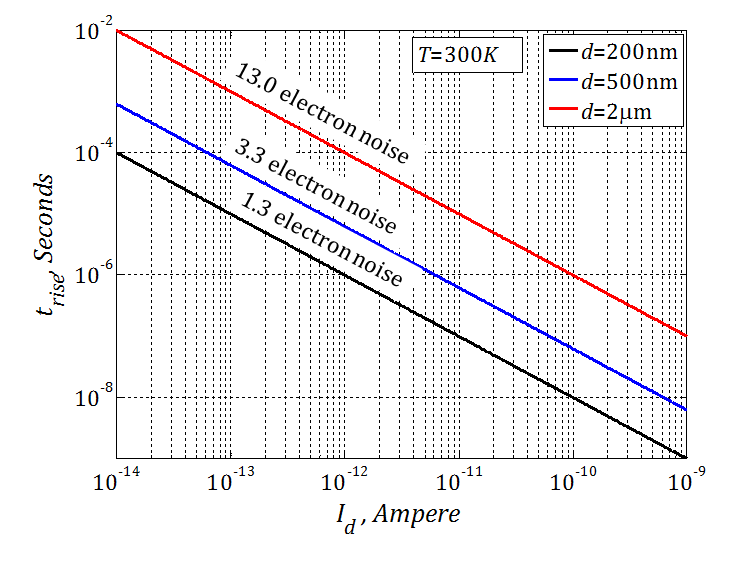}
\caption{\label{fig:rise_time} Rise time of an  InP/InGaAs/InGaAs heterojunction PTD versus its internal dark current for different diameters of the junction area (see Ref. \onlinecite{InGaAs_repeat}).
The total amount of the noise per read time is shown inside the figure for each diameter in the units of electrons RMS (root mean square).}
\end{figure}
    Fig. \ref{fig:rise_time}  shows the rise time versus the  internal dark current for the described PTD with different diameters of the electronic area, $d$ \cite{InGaAs_repeat}. Changing  the temperature changes the internal dark current, hence it  changes the speed of the PTD. For each PTD the number of the electrons RMS noise is also shown in the figure.  This number is calculated using the first term in the square root from Eq.~\ref{eq:SNR_simple}. \

\section{Conclusion}
In conclusion, we show that the minimum number of detectable photons for a PTD is independent of its dark current, and is weakly dependent on its temperature. Our modeling shows that at a given temperature, the sensitivity mainly depends on the junction capacitance. Changing the PTD's temperature may change its dark current level. Variation in the dark current level results in variation in the speed of the detection. The only assumption in the proposed modeling is that the read-out noise is canceled by the PTD's internal gain.\ 
\begin{acknowledgments}
The authors acknowledge the  financial support of the W. M. Keck foundation  for a research grant in science and engineering.
\end{acknowledgments}
\nocite{*}
\bibliography{aipsamp}

\providecommand{\noopsort}[1]{}\providecommand{\singleletter}[1]{#1}%
\begin{thebibliography}{28}%
\makeatletter
\providecommand \@ifxundefined [1]{%
 \@ifx{#1\undefined}
}%
\providecommand \@ifnum [1]{%
 \ifnum #1\expandafter \@firstoftwo
 \else \expandafter \@secondoftwo
 \fi
}%
\providecommand \@ifx [1]{%
 \ifx #1\expandafter \@firstoftwo
 \else \expandafter \@secondoftwo
 \fi
}%
\providecommand \natexlab [1]{#1}%
\providecommand \enquote  [1]{``#1''}%
\providecommand \bibnamefont  [1]{#1}%
\providecommand \bibfnamefont [1]{#1}%
\providecommand \citenamefont [1]{#1}%
\providecommand \href@noop [0]{\@secondoftwo}%
\providecommand \href [0]{\begingroup \@sanitize@url \@href}%
\providecommand \@href[1]{\@@startlink{#1}\@@href}%
\providecommand \@@href[1]{\endgroup#1\@@endlink}%
\providecommand \@sanitize@url [0]{\catcode `\\12\catcode `\$12\catcode
  `\&12\catcode `\#12\catcode `\^12\catcode `\_12\catcode `\%12\relax}%
\providecommand \@@startlink[1]{}%
\providecommand \@@endlink[0]{}%
\providecommand \url  [0]{\begingroup\@sanitize@url \@url }%
\providecommand \@url [1]{\endgroup\@href {#1}{\urlprefix }}%
\providecommand \urlprefix  [0]{URL }%
\providecommand \Eprint [0]{\href }%
\providecommand \doibase [0]{http://dx.doi.org/}%
\providecommand \selectlanguage [0]{\@gobble}%
\providecommand \bibinfo  [0]{\@secondoftwo}%
\providecommand \bibfield  [0]{\@secondoftwo}%
\providecommand \translation [1]{[#1]}%
\providecommand \BibitemOpen [0]{}%
\providecommand \bibitemStop [0]{}%
\providecommand \bibitemNoStop [0]{.\EOS\space}%
\providecommand \EOS [0]{\spacefactor3000\relax}%
\providecommand \BibitemShut  [1]{\csname bibitem#1\endcsname}%
\let\auto@bib@innerbib\@empty
\bibitem [{\citenamefont {Zhang}\ \emph {et~al.}(2015)\citenamefont {Zhang},
  \citenamefont {Itzler}, \citenamefont {Zbinden},\ and\ \citenamefont
  {Pan}}]{Review_SPD}%
  \BibitemOpen
  \bibfield  {author} {\bibinfo {author} {\bibfnamefont {J.}~\bibnamefont
  {Zhang}}, \bibinfo {author} {\bibfnamefont {M.~A.}\ \bibnamefont {Itzler}},
  \bibinfo {author} {\bibfnamefont {H.}~\bibnamefont {Zbinden}}, \ and\
  \bibinfo {author} {\bibfnamefont {J.~W.}\ \bibnamefont {Pan}},\ }\bibfield
  {title} {\enquote {\bibinfo {title} {Advances in ingaas/inp single-photon
  detector systems for quantum communication},}\ }\href {\doibase
  doi:10.1038/lsa.2015.59} {\bibfield  {journal} {\bibinfo  {journal} {Light:
  Science and Applications}\ }\textbf {\bibinfo {volume} {4}},\ \bibinfo
  {pages} {e286} (\bibinfo {year} {2015})}\BibitemShut {NoStop}%
\bibitem [{\citenamefont {Weng}\ \emph {et~al.}(2011)\citenamefont {Weng},
  \citenamefont {Chang}, \citenamefont {Hsu},\ and\ \citenamefont
  {Hsueh}}]{Nano-Wire_PT1}%
  \BibitemOpen
  \bibfield  {author} {\bibinfo {author} {\bibfnamefont {W.~Y.}\ \bibnamefont
  {Weng}}, \bibinfo {author} {\bibfnamefont {S.~J.}\ \bibnamefont {Chang}},
  \bibinfo {author} {\bibfnamefont {C.~L.}\ \bibnamefont {Hsu}}, \ and\
  \bibinfo {author} {\bibfnamefont {T.~J.}\ \bibnamefont {Hsueh}},\ }\bibfield
  {title} {\enquote {\bibinfo {title} {A zno-nanowire phototransistor prepared
  on glass substrates},}\ }\href {\doibase 10.1021/am100746c} {\bibfield
  {journal} {\bibinfo  {journal} {ACS Applied Materials \& Interfaces}\
  }\textbf {\bibinfo {volume} {3}},\ \bibinfo {pages} {162--166} (\bibinfo
  {year} {2011})},\ \bibinfo {note} {pMID: 21226533},\ \Eprint
  {http://arxiv.org/abs/http://dx.doi.org/10.1021/am100746c}
  {http://dx.doi.org/10.1021/am100746c} \BibitemShut {NoStop}%
\bibitem [{\citenamefont {Zhang}\ \emph {et~al.}(2010)\citenamefont {Zhang},
  \citenamefont {Kim}, \citenamefont {Cheng},\ and\ \citenamefont
  {Lo}}]{Nano_wire_PT2}%
  \BibitemOpen
  \bibfield  {author} {\bibinfo {author} {\bibfnamefont {A.}~\bibnamefont
  {Zhang}}, \bibinfo {author} {\bibfnamefont {H.}~\bibnamefont {Kim}}, \bibinfo
  {author} {\bibfnamefont {J.}~\bibnamefont {Cheng}}, \ and\ \bibinfo {author}
  {\bibfnamefont {Y.-H.}\ \bibnamefont {Lo}},\ }\bibfield  {title} {\enquote
  {\bibinfo {title} {Ultrahigh responsivity visible and infrared detection
  using silicon nanowire phototransistors},}\ }\href {\doibase
  10.1021/nl1006432} {\bibfield  {journal} {\bibinfo  {journal} {Nano Letters}\
  }\textbf {\bibinfo {volume} {10}},\ \bibinfo {pages} {2117--2120} (\bibinfo
  {year} {2010})},\ \bibinfo {note} {pMID: 20469840},\ \Eprint
  {http://arxiv.org/abs/http://dx.doi.org/10.1021/nl1006432}
  {http://dx.doi.org/10.1021/nl1006432} \BibitemShut {NoStop}%
\bibitem [{\citenamefont {Yan}, \citenamefont {Gargas},\ and\ \citenamefont
  {Yang}(2009)}]{Nano1}%
  \BibitemOpen
  \bibfield  {author} {\bibinfo {author} {\bibfnamefont {R.}~\bibnamefont
  {Yan}}, \bibinfo {author} {\bibfnamefont {D.}~\bibnamefont {Gargas}}, \ and\
  \bibinfo {author} {\bibfnamefont {P.}~\bibnamefont {Yang}},\ }\bibfield
  {title} {\enquote {\bibinfo {title} {Nanowire photonics},}\ }\href {\doibase
  doi:10.1038/nphoton.2009.184} {\bibfield  {journal} {\bibinfo  {journal}
  {Nature Photonics}\ }\textbf {\bibinfo {volume} {3}},\ \bibinfo {pages}
  {659--576} (\bibinfo {year} {2009})}\BibitemShut {NoStop}%
\bibitem [{\citenamefont {Soci}\ \emph {et~al.}(2007)\citenamefont {Soci},
  \citenamefont {Zhang}, \citenamefont {Xiang}, \citenamefont {Dayeh},
  \citenamefont {Aplin}, \citenamefont {Park}, \citenamefont {Bao},
  \citenamefont {Lo},\ and\ \citenamefont {Wang}}]{nano_ZNO}%
  \BibitemOpen
  \bibfield  {author} {\bibinfo {author} {\bibfnamefont {C.}~\bibnamefont
  {Soci}}, \bibinfo {author} {\bibfnamefont {A.}~\bibnamefont {Zhang}},
  \bibinfo {author} {\bibfnamefont {B.}~\bibnamefont {Xiang}}, \bibinfo
  {author} {\bibfnamefont {S.~A.}\ \bibnamefont {Dayeh}}, \bibinfo {author}
  {\bibfnamefont {D.~P.~R.}\ \bibnamefont {Aplin}}, \bibinfo {author}
  {\bibfnamefont {J.}~\bibnamefont {Park}}, \bibinfo {author} {\bibfnamefont
  {X.~Y.}\ \bibnamefont {Bao}}, \bibinfo {author} {\bibfnamefont {Y.~H.}\
  \bibnamefont {Lo}}, \ and\ \bibinfo {author} {\bibfnamefont {D.}~\bibnamefont
  {Wang}},\ }\bibfield  {title} {\enquote {\bibinfo {title} {Zno nanowire uv
  photodetectors with high internal gain},}\ }\href {\doibase
  10.1021/nl070111x} {\bibfield  {journal} {\bibinfo  {journal} {Nano Letters}\
  }\textbf {\bibinfo {volume} {7}},\ \bibinfo {pages} {1003--1009} (\bibinfo
  {year} {2007})},\ \bibinfo {note} {pMID: 17358092},\ \Eprint
  {http://arxiv.org/abs/http://dx.doi.org/10.1021/nl070111x}
  {http://dx.doi.org/10.1021/nl070111x} \BibitemShut {NoStop}%
\bibitem [{\citenamefont {Hayden}, \citenamefont {Agarwal},\ and\ \citenamefont
  {M.~Lieber}(2006)}]{nture_APD}%
  \BibitemOpen
  \bibfield  {author} {\bibinfo {author} {\bibfnamefont {O.}~\bibnamefont
  {Hayden}}, \bibinfo {author} {\bibfnamefont {R.}~\bibnamefont {Agarwal}}, \
  and\ \bibinfo {author} {\bibfnamefont {C.}~\bibnamefont {M.~Lieber}},\
  }\bibfield  {title} {\enquote {\bibinfo {title} {Nanoscale avalanche
  photodiodes for highly sensitive and spatially resolved photon detection},}\
  }\href {\doibase 10.1038/nmat1635} {\bibfield  {journal} {\bibinfo  {journal}
  {Nature Materials}\ }\textbf {\bibinfo {volume} {5}},\ \bibinfo {pages}
  {352--356} (\bibinfo {year} {2006})}\BibitemShut {NoStop}%
\bibitem [{\citenamefont {Huang}\ \emph {et~al.}(2002)\citenamefont {Huang},
  \citenamefont {Duan}, \citenamefont {Cui},\ and\ \citenamefont
  {Lieber}}]{Nano_GAN}%
  \BibitemOpen
  \bibfield  {author} {\bibinfo {author} {\bibfnamefont {Y.}~\bibnamefont
  {Huang}}, \bibinfo {author} {\bibfnamefont {X.}~\bibnamefont {Duan}},
  \bibinfo {author} {\bibfnamefont {Y.}~\bibnamefont {Cui}}, \ and\ \bibinfo
  {author} {\bibfnamefont {C.~M.}\ \bibnamefont {Lieber}},\ }\bibfield  {title}
  {\enquote {\bibinfo {title} {Gallium nitride nanowire nanodevices},}\ }\href
  {\doibase 10.1021/nl015667d} {\bibfield  {journal} {\bibinfo  {journal} {Nano
  Letters}\ }\textbf {\bibinfo {volume} {2}},\ \bibinfo {pages} {101--104}
  (\bibinfo {year} {2002})},\ \Eprint
  {http://arxiv.org/abs/http://dx.doi.org/10.1021/nl015667d}
  {http://dx.doi.org/10.1021/nl015667d} \BibitemShut {NoStop}%
\bibitem [{\citenamefont {Guo}\ \emph {et~al.}(2014)\citenamefont {Guo},
  \citenamefont {Hu}, \citenamefont {Liao}, \citenamefont {Yip}, \citenamefont
  {Ho}, \citenamefont {Miao}, \citenamefont {Zhang}, \citenamefont {Zou},
  \citenamefont {Jiang}, \citenamefont {Wu}, \citenamefont {Chen},\ and\
  \citenamefont {Lu}}]{Nano_PT_InAs}%
  \BibitemOpen
  \bibfield  {author} {\bibinfo {author} {\bibfnamefont {N.}~\bibnamefont
  {Guo}}, \bibinfo {author} {\bibfnamefont {W.}~\bibnamefont {Hu}}, \bibinfo
  {author} {\bibfnamefont {L.}~\bibnamefont {Liao}}, \bibinfo {author}
  {\bibfnamefont {S.}~\bibnamefont {Yip}}, \bibinfo {author} {\bibfnamefont
  {J.~C.}\ \bibnamefont {Ho}}, \bibinfo {author} {\bibfnamefont
  {J.}~\bibnamefont {Miao}}, \bibinfo {author} {\bibfnamefont {Z.}~\bibnamefont
  {Zhang}}, \bibinfo {author} {\bibfnamefont {J.}~\bibnamefont {Zou}}, \bibinfo
  {author} {\bibfnamefont {T.}~\bibnamefont {Jiang}}, \bibinfo {author}
  {\bibfnamefont {S.}~\bibnamefont {Wu}}, \bibinfo {author} {\bibfnamefont
  {X.}~\bibnamefont {Chen}}, \ and\ \bibinfo {author} {\bibfnamefont
  {W.}~\bibnamefont {Lu}},\ }\bibfield  {title} {\enquote {\bibinfo {title}
  {Nanowires: Anomalous and highly efficient inas nanowire phototransistors
  based on majority carrier transport at room temperature (adv. mater.
  48/2014)},}\ }\href {\doibase 10.1002/adma.201470329} {\bibfield  {journal}
  {\bibinfo  {journal} {Advanced Materials}\ }\textbf {\bibinfo {volume}
  {26}},\ \bibinfo {pages} {8232--8232} (\bibinfo {year} {2014})}\BibitemShut
  {NoStop}%
\bibitem [{\citenamefont {Tan}\ \emph {et~al.}(2016)\citenamefont {Tan},
  \citenamefont {Zhao}, \citenamefont {Chen}, \citenamefont {Crozier},\ and\
  \citenamefont {Dan}}]{Silicon_nano}%
  \BibitemOpen
  \bibfield  {author} {\bibinfo {author} {\bibfnamefont {S.~L.}\ \bibnamefont
  {Tan}}, \bibinfo {author} {\bibfnamefont {X.}~\bibnamefont {Zhao}}, \bibinfo
  {author} {\bibfnamefont {K.}~\bibnamefont {Chen}}, \bibinfo {author}
  {\bibfnamefont {K.~B.}\ \bibnamefont {Crozier}}, \ and\ \bibinfo {author}
  {\bibfnamefont {Y.}~\bibnamefont {Dan}},\ }\bibfield  {title} {\enquote
  {\bibinfo {title} {High-performance silicon nanowire bipolar
  phototransistors},}\ }\href {\doibase 10.1063/1.4959264} {\bibfield
  {journal} {\bibinfo  {journal} {Applied Physics Letters}\ }\textbf {\bibinfo
  {volume} {109}},\ \bibinfo {pages} {033505} (\bibinfo {year} {2016})},\
  \Eprint {http://arxiv.org/abs/http://dx.doi.org/10.1063/1.4959264}
  {http://dx.doi.org/10.1063/1.4959264} \BibitemShut {NoStop}%
\bibitem [{\citenamefont {Movassaghi}\ \emph {et~al.}(2016)\citenamefont
  {Movassaghi}, \citenamefont {Fathipour}, \citenamefont {Fathipour},\ and\
  \citenamefont {Mohseni}}]{yashar}%
  \BibitemOpen
  \bibfield  {author} {\bibinfo {author} {\bibfnamefont {Y.}~\bibnamefont
  {Movassaghi}}, \bibinfo {author} {\bibfnamefont {V.}~\bibnamefont
  {Fathipour}}, \bibinfo {author} {\bibfnamefont {M.}~\bibnamefont
  {Fathipour}}, \ and\ \bibinfo {author} {\bibfnamefont {H.}~\bibnamefont
  {Mohseni}},\ }\bibfield  {title} {\enquote {\bibinfo {title} {Analytical
  modeling and numerical simulation of the short-wave infrared
  electron-injection detectors},}\ }\href {\doibase 10.1063/1.4944602}
  {\bibfield  {journal} {\bibinfo  {journal} {Applied Physics Letters}\
  }\textbf {\bibinfo {volume} {108}},\ \bibinfo {pages} {121102} (\bibinfo
  {year} {2016})},\ \Eprint
  {http://arxiv.org/abs/http://dx.doi.org/10.1063/1.4944602}
  {http://dx.doi.org/10.1063/1.4944602} \BibitemShut {NoStop}%
\bibitem [{\citenamefont {Namekata}, \citenamefont {Sasamori},\ and\
  \citenamefont {Inoue}(2006)}]{avalanche}%
  \BibitemOpen
  \bibfield  {author} {\bibinfo {author} {\bibfnamefont {N.}~\bibnamefont
  {Namekata}}, \bibinfo {author} {\bibfnamefont {S.}~\bibnamefont {Sasamori}},
  \ and\ \bibinfo {author} {\bibfnamefont {S.}~\bibnamefont {Inoue}},\
  }\bibfield  {title} {\enquote {\bibinfo {title} {800 mhz single-photon
  detection at 1550-nm using an ingaas/inp avalanche photodiode operated with a
  sine wave gating},}\ }\href {\doibase 10.1364/OE.14.010043} {\bibfield
  {journal} {\bibinfo  {journal} {Opt. Express}\ }\textbf {\bibinfo {volume}
  {14}},\ \bibinfo {pages} {10043--10049} (\bibinfo {year} {2006})}\BibitemShut
  {NoStop}%
\bibitem [{\citenamefont {Leu}, \citenamefont {Gardner},\ and\ \citenamefont
  {Forrest}(1991)}]{opt_communication}%
  \BibitemOpen
  \bibfield  {author} {\bibinfo {author} {\bibfnamefont {L.~Y.}\ \bibnamefont
  {Leu}}, \bibinfo {author} {\bibfnamefont {J.~T.}\ \bibnamefont {Gardner}}, \
  and\ \bibinfo {author} {\bibfnamefont {S.~R.}\ \bibnamefont {Forrest}},\
  }\bibfield  {title} {\enquote {\bibinfo {title} {A high‐gain,
  high‐bandwidth in0.53ga0.47as/inp heterojunction phototransistor for
  optical communications},}\ }\href {\doibase 10.1063/1.347371} {\bibfield
  {journal} {\bibinfo  {journal} {Journal of Applied Physics}\ }\textbf
  {\bibinfo {volume} {69}},\ \bibinfo {pages} {1052--1062} (\bibinfo {year}
  {1991})},\ \Eprint {http://arxiv.org/abs/http://dx.doi.org/10.1063/1.347371}
  {http://dx.doi.org/10.1063/1.347371} \BibitemShut {NoStop}%
\bibitem [{\citenamefont {Memis}\ \emph {et~al.}(2007)\citenamefont {Memis},
  \citenamefont {Katsnelson}, \citenamefont {Kong}, \citenamefont {Mohseni},
  \citenamefont {Yan}, \citenamefont {Zhang}, \citenamefont {Hossain},
  \citenamefont {Jin},\ and\ \citenamefont {Adesida}}]{Gokab-main}%
  \BibitemOpen
  \bibfield  {author} {\bibinfo {author} {\bibfnamefont {O.~G.}\ \bibnamefont
  {Memis}}, \bibinfo {author} {\bibfnamefont {A.}~\bibnamefont {Katsnelson}},
  \bibinfo {author} {\bibfnamefont {S.-C.}\ \bibnamefont {Kong}}, \bibinfo
  {author} {\bibfnamefont {H.}~\bibnamefont {Mohseni}}, \bibinfo {author}
  {\bibfnamefont {M.}~\bibnamefont {Yan}}, \bibinfo {author} {\bibfnamefont
  {S.}~\bibnamefont {Zhang}}, \bibinfo {author} {\bibfnamefont
  {T.}~\bibnamefont {Hossain}}, \bibinfo {author} {\bibfnamefont
  {N.}~\bibnamefont {Jin}}, \ and\ \bibinfo {author} {\bibfnamefont
  {I.}~\bibnamefont {Adesida}},\ }\bibfield  {title} {\enquote {\bibinfo
  {title} {A photon detector with very high gain at low bias and at room
  temperature},}\ }\href {\doibase 10.1063/1.2802043} {\bibfield  {journal}
  {\bibinfo  {journal} {Applied Physics Letters}\ }\textbf {\bibinfo {volume}
  {91}},\ \bibinfo {pages} {171112} (\bibinfo {year} {2007})},\ \Eprint
  {http://arxiv.org/abs/http://dx.doi.org/10.1063/1.2802043}
  {http://dx.doi.org/10.1063/1.2802043} \BibitemShut {NoStop}%
\bibitem [{\citenamefont {Dehzangi}\ \emph {et~al.}(2017)\citenamefont
  {Dehzangi}, \citenamefont {Haddadi}, \citenamefont {Adhikary},\ and\
  \citenamefont {Razeghi}}]{Arash}%
  \BibitemOpen
  \bibfield  {author} {\bibinfo {author} {\bibfnamefont {A.}~\bibnamefont
  {Dehzangi}}, \bibinfo {author} {\bibfnamefont {A.}~\bibnamefont {Haddadi}},
  \bibinfo {author} {\bibfnamefont {S.}~\bibnamefont {Adhikary}}, \ and\
  \bibinfo {author} {\bibfnamefont {M.}~\bibnamefont {Razeghi}},\ }\bibfield
  {title} {\enquote {\bibinfo {title} {Impact of scaling base thickness on the
  performance of heterojunction phototransistors},}\ }\href
  {http://stacks.iop.org/0957-4484/28/i=10/a=10LT01} {\bibfield  {journal}
  {\bibinfo  {journal} {Nanotechnology}\ }\textbf {\bibinfo {volume} {28}},\
  \bibinfo {pages} {10LT01} (\bibinfo {year} {2017})}\BibitemShut {NoStop}%
\bibitem [{\citenamefont {Park}\ and\ \citenamefont {Jang}(2010)}]{Min_Su}%
  \BibitemOpen
  \bibfield  {author} {\bibinfo {author} {\bibfnamefont {M.~S.}\ \bibnamefont
  {Park}}\ and\ \bibinfo {author} {\bibfnamefont {J.~H.}\ \bibnamefont
  {Jang}},\ }\bibfield  {title} {\enquote {\bibinfo {title} {Enhancement of
  optical gain in floating-base ingap-gaas heterojunction phototransistors},}\
  }\href {\doibase 10.1109/LPT.2010.2051660} {\bibfield  {journal} {\bibinfo
  {journal} {IEEE Photonics Technology Letters}\ }\textbf {\bibinfo {volume}
  {22}},\ \bibinfo {pages} {1202--1204} (\bibinfo {year} {2010})}\BibitemShut
  {NoStop}%
\bibitem [{\citenamefont {Rezaei}\ \emph {et~al.}(2016)\citenamefont {Rezaei},
  \citenamefont {Park}, \citenamefont {Wheaton}, \citenamefont {Tan},
  \citenamefont {Fathipour}, \citenamefont {Guyon}, \citenamefont {Ulmer},\
  and\ \citenamefont {Mohseni}}]{mohsen}%
  \BibitemOpen
  \bibfield  {author} {\bibinfo {author} {\bibfnamefont {M.}~\bibnamefont
  {Rezaei}}, \bibinfo {author} {\bibfnamefont {M.-S.}\ \bibnamefont {Park}},
  \bibinfo {author} {\bibfnamefont {S.}~\bibnamefont {Wheaton}}, \bibinfo
  {author} {\bibfnamefont {C.~L.}\ \bibnamefont {Tan}}, \bibinfo {author}
  {\bibfnamefont {V.}~\bibnamefont {Fathipour}}, \bibinfo {author}
  {\bibfnamefont {O.}~\bibnamefont {Guyon}}, \bibinfo {author} {\bibfnamefont
  {M.~P.}\ \bibnamefont {Ulmer}}, \ and\ \bibinfo {author} {\bibfnamefont
  {H.}~\bibnamefont {Mohseni}},\ }\bibfield  {title} {\enquote {\bibinfo
  {title} {New progress in electron-injection detectors for nir imagers with
  low noise and high frame rates},}\ }\href {\doibase 10.1117/12.2233657}
  {\bibfield  {journal} {\bibinfo  {journal} {Proc. SPIE}\ }\textbf {\bibinfo
  {volume} {9915}},\ \bibinfo {pages} {99150P--99150P--8} (\bibinfo {year}
  {2016})}\BibitemShut {NoStop}%
\bibitem [{\citenamefont {Helme}\ and\ \citenamefont
  {Houston}(2007)}]{PT-analytic}%
  \BibitemOpen
  \bibfield  {author} {\bibinfo {author} {\bibfnamefont {J.~P.}\ \bibnamefont
  {Helme}}\ and\ \bibinfo {author} {\bibfnamefont {P.~A.}\ \bibnamefont
  {Houston}},\ }\bibfield  {title} {\enquote {\bibinfo {title} {Analytical
  modeling of speed response of heterojunction bipolar phototransistors},}\
  }\href {\doibase 10.1109/JLT.2007.893891} {\bibfield  {journal} {\bibinfo
  {journal} {Journal of Lightwave Technology}\ }\textbf {\bibinfo {volume}
  {25}},\ \bibinfo {pages} {1247--1255} (\bibinfo {year} {2007})}\BibitemShut
  {NoStop}%
\bibitem [{\citenamefont {Gaberl}\ \emph {et~al.}(2014)\citenamefont {Gaberl},
  \citenamefont {Kostov}, \citenamefont {Hofbauer},\ and\ \citenamefont
  {Zimmermann}}]{Noise-PT_2014}%
  \BibitemOpen
  \bibfield  {author} {\bibinfo {author} {\bibfnamefont {W.}~\bibnamefont
  {Gaberl}}, \bibinfo {author} {\bibfnamefont {P.}~\bibnamefont {Kostov}},
  \bibinfo {author} {\bibfnamefont {M.}~\bibnamefont {Hofbauer}}, \ and\
  \bibinfo {author} {\bibfnamefont {H.}~\bibnamefont {Zimmermann}},\ }\bibfield
   {title} {\enquote {\bibinfo {title} {Phototransistor noise model based on
  noise measurements on pnp pin phototransistors},}\ }\href {\doibase
  10.1007/s11082-013-9839-1} {\bibfield  {journal} {\bibinfo  {journal}
  {Optical and Quantum Electronics}\ }\textbf {\bibinfo {volume} {46}},\
  \bibinfo {pages} {1269--1275} (\bibinfo {year} {2014})}\BibitemShut {NoStop}%
\bibitem [{\citenamefont {Memis}\ \emph {et~al.}(2008)\citenamefont {Memis},
  \citenamefont {Katsnelson}, \citenamefont {Kong}, \citenamefont {Mohseni},
  \citenamefont {Yan}, \citenamefont {Zhang}, \citenamefont {Hossain},
  \citenamefont {Jin},\ and\ \citenamefont {Adesida}}]{Memis-noise}%
  \BibitemOpen
  \bibfield  {author} {\bibinfo {author} {\bibfnamefont {O.~G.}\ \bibnamefont
  {Memis}}, \bibinfo {author} {\bibfnamefont {A.}~\bibnamefont {Katsnelson}},
  \bibinfo {author} {\bibfnamefont {S.-C.}\ \bibnamefont {Kong}}, \bibinfo
  {author} {\bibfnamefont {H.}~\bibnamefont {Mohseni}}, \bibinfo {author}
  {\bibfnamefont {M.}~\bibnamefont {Yan}}, \bibinfo {author} {\bibfnamefont
  {S.}~\bibnamefont {Zhang}}, \bibinfo {author} {\bibfnamefont
  {T.}~\bibnamefont {Hossain}}, \bibinfo {author} {\bibfnamefont
  {N.}~\bibnamefont {Jin}}, \ and\ \bibinfo {author} {\bibfnamefont
  {I.}~\bibnamefont {Adesida}},\ }\bibfield  {title} {\enquote {\bibinfo
  {title} {Sub-poissonian shot noise of a high internal gain injection photon
  detector},}\ }\href {\doibase 10.1364/OE.16.012701} {\bibfield  {journal}
  {\bibinfo  {journal} {Opt. Express}\ }\textbf {\bibinfo {volume} {16}},\
  \bibinfo {pages} {12701--12706} (\bibinfo {year} {2008})}\BibitemShut
  {NoStop}%
\bibitem [{\citenamefont {Blanter}\ and\ \citenamefont
  {Büttiker}(2000)}]{Shot-noise-Review}%
  \BibitemOpen
  \bibfield  {author} {\bibinfo {author} {\bibfnamefont {Y.}~\bibnamefont
  {Blanter}}\ and\ \bibinfo {author} {\bibfnamefont {M.}~\bibnamefont
  {Büttiker}},\ }\bibfield  {title} {\enquote {\bibinfo {title} {Shot noise in
  mesoscopic conductors},}\ }\href {\doibase
  http://dx.doi.org/10.1016/S0370-1573(99)00123-4} {\bibfield  {journal}
  {\bibinfo  {journal} {Physics Reports}\ }\textbf {\bibinfo {volume} {336}},\
  \bibinfo {pages} {1 -- 166} (\bibinfo {year} {2000})}\BibitemShut {NoStop}%
\bibitem [{\citenamefont {Spieler}(2005)}]{book}%
  \BibitemOpen
  \bibfield  {author} {\bibinfo {author} {\bibfnamefont {H.}~\bibnamefont
  {Spieler}},\ }\href@noop {} {\emph {\bibinfo {title} {Semiconductor detector
  systems}}},\ Vol.~\bibinfo {volume} {12}\ (\bibinfo  {publisher} {Oxford
  university press},\ \bibinfo {year} {2005})\BibitemShut {NoStop}%
\bibitem [{\citenamefont {Chand}, \citenamefont {Houston},\ and\ \citenamefont
  {Robson}(1985)}]{gain_of_HPT}%
  \BibitemOpen
  \bibfield  {author} {\bibinfo {author} {\bibfnamefont {N.}~\bibnamefont
  {Chand}}, \bibinfo {author} {\bibfnamefont {P.~A.}\ \bibnamefont {Houston}},
  \ and\ \bibinfo {author} {\bibfnamefont {P.~N.}\ \bibnamefont {Robson}},\
  }\bibfield  {title} {\enquote {\bibinfo {title} {Gain of a heterojunction
  bipolar phototransistor},}\ }\href {\doibase 10.1109/T-ED.1985.21988}
  {\bibfield  {journal} {\bibinfo  {journal} {IEEE Transactions on Electron
  Devices}\ }\textbf {\bibinfo {volume} {32}},\ \bibinfo {pages} {622--627}
  (\bibinfo {year} {1985})}\BibitemShut {NoStop}%
\bibitem [{\citenamefont {Choi}\ \emph {et~al.}(2009)\citenamefont {Choi},
  \citenamefont {Furue}, \citenamefont {Hayama}, \citenamefont {Nishida},\ and\
  \citenamefont {Ogura}}]{InGaAs_repeat}%
  \BibitemOpen
  \bibfield  {author} {\bibinfo {author} {\bibfnamefont {S.~W.}\ \bibnamefont
  {Choi}}, \bibinfo {author} {\bibfnamefont {S.}~\bibnamefont {Furue}},
  \bibinfo {author} {\bibfnamefont {N.}~\bibnamefont {Hayama}}, \bibinfo
  {author} {\bibfnamefont {K.}~\bibnamefont {Nishida}}, \ and\ \bibinfo
  {author} {\bibfnamefont {M.}~\bibnamefont {Ogura}},\ }\bibfield  {title}
  {\enquote {\bibinfo {title} {Gain-enhanced ingaas-inp heterojunction
  phototransistor with zn-doped mesa sidewall},}\ }\href {\doibase
  10.1109/LPT.2009.2023367} {\bibfield  {journal} {\bibinfo  {journal} {IEEE
  Photonics Technology Letters}\ }\textbf {\bibinfo {volume} {21}},\ \bibinfo
  {pages} {1187--1189} (\bibinfo {year} {2009})}\BibitemShut {NoStop}%
\bibitem [{\citenamefont {Moneda}, \citenamefont {Chenette},\ and\
  \citenamefont {Ziel}(1971)}]{Noise-PT_1971}%
  \BibitemOpen
  \bibfield  {author} {\bibinfo {author} {\bibfnamefont {F.~H. D.~L.}\
  \bibnamefont {Moneda}}, \bibinfo {author} {\bibfnamefont {E.~R.}\
  \bibnamefont {Chenette}}, \ and\ \bibinfo {author} {\bibfnamefont {A.~V.~D.}\
  \bibnamefont {Ziel}},\ }\bibfield  {title} {\enquote {\bibinfo {title} {Noise
  in phototransistors},}\ }\href {\doibase 10.1109/T-ED.1971.17198} {\bibfield
  {journal} {\bibinfo  {journal} {IEEE Transactions on Electron Devices}\
  }\textbf {\bibinfo {volume} {18}},\ \bibinfo {pages} {340--346} (\bibinfo
  {year} {1971})}\BibitemShut {NoStop}%
\bibitem [{\citenamefont {Campbell}\ and\ \citenamefont
  {Ogawa}(1982)}]{Optical_receiver_1982}%
  \BibitemOpen
  \bibfield  {author} {\bibinfo {author} {\bibfnamefont {J.~C.}\ \bibnamefont
  {Campbell}}\ and\ \bibinfo {author} {\bibfnamefont {K.}~\bibnamefont
  {Ogawa}},\ }\bibfield  {title} {\enquote {\bibinfo {title} {Heterojunction
  phototransistors for long‐wavelength optical receivers},}\ }\href@noop {}
  {\bibfield  {journal} {\bibinfo  {journal} {Journal of Applied Physics}\
  }\textbf {\bibinfo {volume} {53}} (\bibinfo {year} {1982})}\BibitemShut
  {NoStop}%
\bibitem [{\citenamefont {Ahrenkiel}, \citenamefont {Keyes},\ and\
  \citenamefont {Dunlavy}(1991)}]{power-dependent-lifetime}%
  \BibitemOpen
  \bibfield  {author} {\bibinfo {author} {\bibfnamefont {R.~K.}\ \bibnamefont
  {Ahrenkiel}}, \bibinfo {author} {\bibfnamefont {B.~M.}\ \bibnamefont
  {Keyes}}, \ and\ \bibinfo {author} {\bibfnamefont {D.~J.}\ \bibnamefont
  {Dunlavy}},\ }\bibfield  {title} {\enquote {\bibinfo {title}
  {Intensity‐dependent minority‐carrier lifetime in iii‐v semiconductors
  due to saturation of recombination centers},}\ }\href@noop {} {\bibfield
  {journal} {\bibinfo  {journal} {Journal of Applied Physics}\ }\textbf
  {\bibinfo {volume} {70}} (\bibinfo {year} {1991})}\BibitemShut {NoStop}%
\bibitem [{\citenamefont {Campbell}\ \emph {et~al.}(1981)\citenamefont
  {Campbell}, \citenamefont {Dentai}, \citenamefont {Burrus},\ and\
  \citenamefont {Ferguson}}]{Optical_fiber_receivers}%
  \BibitemOpen
  \bibfield  {author} {\bibinfo {author} {\bibfnamefont {J.}~\bibnamefont
  {Campbell}}, \bibinfo {author} {\bibfnamefont {A.}~\bibnamefont {Dentai}},
  \bibinfo {author} {\bibfnamefont {C.}~\bibnamefont {Burrus}}, \ and\ \bibinfo
  {author} {\bibfnamefont {J.}~\bibnamefont {Ferguson}},\ }\bibfield  {title}
  {\enquote {\bibinfo {title} {Inp/ingaas heterojunction phototransistors},}\
  }\href {\doibase 10.1109/JQE.1981.1071072} {\bibfield  {journal} {\bibinfo
  {journal} {IEEE Journal of Quantum Electronics}\ }\textbf {\bibinfo {volume}
  {17}},\ \bibinfo {pages} {264--269} (\bibinfo {year} {1981})}\BibitemShut
  {NoStop}%
\bibitem [{\citenamefont {Ritter}\ \emph {et~al.}(1991)\citenamefont {Ritter},
  \citenamefont {Hamm}, \citenamefont {Feygenson}, \citenamefont {Panish},\
  and\ \citenamefont {Chandrasekhar}}]{Gain-HBT-1991}%
  \BibitemOpen
  \bibfield  {author} {\bibinfo {author} {\bibfnamefont {D.}~\bibnamefont
  {Ritter}}, \bibinfo {author} {\bibfnamefont {R.~A.}\ \bibnamefont {Hamm}},
  \bibinfo {author} {\bibfnamefont {A.}~\bibnamefont {Feygenson}}, \bibinfo
  {author} {\bibfnamefont {M.~B.}\ \bibnamefont {Panish}}, \ and\ \bibinfo
  {author} {\bibfnamefont {S.}~\bibnamefont {Chandrasekhar}},\ }\bibfield
  {title} {\enquote {\bibinfo {title} {Diffusive base transport in narrow base
  inp/ga0.47in0.53as heterojunction bipolar transistors},}\ }\href@noop {}
  {\bibfield  {journal} {\bibinfo  {journal} {Applied Physics Letters}\
  }\textbf {\bibinfo {volume} {59}} (\bibinfo {year} {1991})}\BibitemShut
  {NoStop}%
\end{thebibliography}%
\end{document}